\documentclass[preprint,12pt]{elsarticle}

\AtBeginDocument{%
  \providecommand\BibTeX{{%
    \normalfont B\kern-0.5em{\scshape i\kern-0.25em b}\kern-0.8em\TeX}}}
\usepackage{verbatim}
\usepackage{amsmath}
\usepackage{algorithm}

\makeatletter
\def\ps@pprintTitle{%
   \let\@oddhead\@empty
   \let\@evenhead\@empty
   \def\@oddfoot{}%
   \let\@evenfoot\@oddfoot}
\makeatother

\newcommand{\Input}[1]{\Statex \textbf{Input:} #1}
\newcommand{\Output}[1]{\Statex \textbf{Output:} #1}

\newcounter{subsubsubsection}[subsubsection]

\usepackage{multicol}
\usepackage[table]{xcolor}

\usepackage{lineno}
\usepackage{lmodern}
\usepackage{color}
\usepackage{listings}
\usepackage{tabularx}
\usepackage{enumerate}
\usepackage{enumitem}
\usepackage{subcaption} 
\usepackage{multirow} 

\usepackage{multirow}

\usepackage{algorithmicx}
\usepackage{algorithm}
\usepackage[noend]{algpseudocode}
\algnewcommand{\Initialize}[1]{%
  \State \textbf{Initialize:}
  \Statex \hspace*{\algorithmicindent}\parbox[t]{.8\linewidth}{\raggedright #1}
}
\algnewcommand\algorithmicforeach{\textbf{for each}}
\algdef{S}[FOR]{ForEach}[1]{\algorithmicforeach\ #1\ \algorithmicdo}

\usepackage{etoolbox}\AtBeginEnvironment{algorithmic}{\fontsize{7}{7}\selectfont}
\algrenewcommand\alglinenumber[1]{\fontsize{7}{7}\selectfont #1:}
\definecolor{Comments}{rgb}{0.00,0.50,0.00}
\definecolor{KeyWords}{rgb}{0.00,0.00,0.13}
\definecolor{Strings}{rgb}{0.60,0.00,0.00}

\lstdefinestyle{psql}
{
  tabsize=3,
  basicstyle=\small\upshape\ttfamily,
  language=SQL,
  morekeywords={PROVENANCE,BASERELATION,INFLUENCE,COPY,ON,TRANSPROV,TRANSSQL,TRANSXML,CONTRIBUTION,COMPLETE,TRANSITIVE,NONTRANSITIVE,EXPLAIN,SQLTEXT,GRAPH,IS,ANNOT,THIS,XSLT,MAPPROV,cxpath,OF,TRANSACTION,SERIALIZABLE,COMMITTED,INSERT,INTO,WITH,SCN,PROV,IMPORT,FOR,JSON,JSON_TABLE,XMLTABLE,DATABASE,SYSTEM,SWITCH,LOGFILE,MEMBER},
  extendedchars=false,
  keywordstyle=\color{blue},
  mathescape=true,
  escapechar=@,
  sensitive=true,
  stringstyle=\color{Strings},%
  string=[b]'
}

\lstdefinestyle{rsl}
{
tabsize=3,
basicstyle=\small\upshape\ttfamily,
language=C,
morekeywords={RULE,LET,CONDITION,RETURN,AND,FOR,INTO,REWRITE,MATCH,WHERE},
extendedchars=false,
keywordstyle=\color{blue},
mathescape=true,
escapechar=@,
sensitive=true
}

\lstdefinestyle{pseudocode}
{
  tabsize=3,
  basicstyle=\small,
  language=c,
  morekeywords={if,else,foreach,case,return,in,or},
  extendedchars=true,
  mathescape=true,
  literate={:=}{{$\gets$}}1 {<=}{{$\leq$}}1 {!=}{{$\neq$}}1 {append}{{$\listconcat$}}1 {calP}{{$\cal P$}}{2},
  keywordstyle=\color{blue},
  escapechar=&,
  numbers=left,
  numberstyle={\color{green}\small\bf}, 
  stepnumber=1, 
  numbersep=5pt,
}

\lstdefinestyle{xmlstyle}
{
  tabsize=3,
  basicstyle=\small,
  language=xml,
  extendedchars=true,
  mathescape=true,
  escapechar=£,
  tagstyle={\color{blue}},
  usekeywordsintag=true,
  morekeywords={alias,name,id},
  keywordstyle={\color{red}}
}

\lstdefinelanguage{json}{
    basicstyle=\footnotesize\ttfamily,
    numbers=left,
    numberstyle=\scriptsize,
    stepnumber=1,
    numbersep=8pt,
    stringstyle=\color{Strings},%
    showstringspaces=false,
    breaklines=true,
    frame=lines,
    string=[b]"
}

\usepackage[utf8]{inputenc}

\usepackage{graphicx}

\usepackage{hyperref}
\usepackage{todonotes}
\usepackage{amsmath, amssymb}
\usepackage{todonotes}
\usepackage{rotating}
\usepackage{caption}
\usepackage{multirow}
\usepackage{stfloats}
\usepackage[utf8]{inputenc}
\usepackage[tableposition=top]{caption}
\usepackage{tabularx}
\usepackage{float}
\usepackage{array}
\usepackage{enumitem}
\usepackage{setspace}
\usepackage{amsmath}
 \usepackage{upquote}
\usepackage[margin=1in]{geometry}
\begin{document}

\begin{frontmatter}

    \title{MemTraceDB: Reconstructing MySQL User Activity Using ActiviTimeTrace Algorithm}

    \author[uno]{Mahfuzul I. Nissan}
    \ead{minissan@uno.edu}
    \date{}
    
    \begin{abstract}
Database audit and transaction logs are fundamental to forensic investigations, but they are vulnerable to tampering by privileged attackers. Malicious insiders or external threats with administrative access can alter, purge, or temporarily disable logging mechanisms, creating significant blind spots and rendering disk-based records unreliable. Memory analysis offers a vital alternative, providing investigators direct access to volatile artifacts that represent a ground-truth source of recent user activity, even when log files have been compromised.

This paper introduces MemTraceDB, a tool that reconstructs user activity timelines by analyzing raw memory snapshots from the MySQL database process. MemTraceDB utilizes a novel algorithm, ActiviTimeTrace, to systematically extract and correlate forensic artifacts such as user connections and executed queries. Through a series of experiments, I demonstrate MemTraceDB's effectiveness and reveal a critical empirical finding: the MySQL query stack has a finite operational capacity of approximately 9,997 queries. This discovery allows me to establish a practical, data-driven formula for determining the optimal frequency for memory snapshot collection, providing a clear, actionable guideline for investigators. The result is a forensically-sound reconstruction of user activity, independent of compromised disk-based logs.
\end{abstract}

    \begin{keyword}
    Memory Forensics, Database Forensics, Digital Forensics, Database Security, Cybersecurity
    \end{keyword}

\end{frontmatter}
\onehalfspacing

\section{Introduction}

In today's digital era, organizations across sectors like healthcare, finance, and e-commerce rely on databases to manage vast volumes of sensitive information. The audit logs from these systems are critical for operational integrity, regulatory compliance with standards like GDPR~\cite{GDPR} and HIPAA~\cite{HIPAA}, and for detecting insider threats~\cite{mandia2003incident}. However, the integrity of these logs cannot be assumed. A privileged attacker—whether a malicious insider or an external threat with escalated credentials—can alter, purge, or temporarily disable logging, creating significant blind spots for forensic investigators. Given that disk-based logs are fundamentally unreliable, a more direct method is needed to validate their contents and uncover hidden activities.

Memory forensics provides this direct method. By analyzing a raw snapshot of a database process, investigators can access a ground-truth source of recent activity, including unencrypted SQL queries and active user sessions. Unlike disk-based logs, these volatile artifacts cannot be easily tampered with by an attacker. Furthermore, memory analysis bypasses challenges that plague other methods; it is immune to system clock manipulation and avoids the computational overhead of decrypting network traffic, offering a more efficient path to evidence.

In this paper, I introduce MemTraceDB, a tool I developed to reconstruct user activity timelines by analyzing raw memory snapshots from the MySQL database process. MemTraceDB is powered by a novel algorithm, ActiviTimeTrace, which systematically extracts and correlates forensic artifacts—specifically SQL DML and DDL commands and user connection data—to generate a forensically-sound timeline. This approach allows an investigator to trace user actions even when on-disk audit logs have been compromised. The major contributions of this work are as follows:

\vspace{-2mm}
\begin{enumerate}
    \item I identify and analyze the forensic artifacts present in MySQL process memory that describe user activity, characterizing their structure and operational lifetime (Section~\ref{sec:overview}).

    \item I present MemTraceDB, a novel tool that automatically extracts these forensic artifacts from MySQL memory snapshots to reconstruct user activity timelines (Section~\ref{sec:artifacts}).

    \item I evaluate MemTraceDB's capabilities through a series of experiments. The evaluation demonstrates the tool's effectiveness and establishes a practical guideline for evidence collection by identifying the finite operational capacity of the MySQL query stack (Section~\ref{sec:experiments}).
\end{enumerate}

\section{Related Work}
\label{sec:relatedwork}

\subsection{Database Memory Forensics}
Foundational digital forensic analysis uses file carving techniques, which reconstruct data without using file system metadata.
The work in \cite{richard2005scalpel, garfinkel2007carving} presented some of the earliest research around file carving performed as a ``dead analysis'' on disk images. 
As the field of digital forensics  matured, memory forensics ``live analysis'' has emerged \cite{case2017memory}. 
An important application for memory forensic investigation is inspecting runtime code to detect malware (e.g., \cite{case2016detecting}). 
Such work requires not only  carving but an extensive analysis of application and kernel data structures.

DBMSes manage their own internal storage separately from the OS and DBMS files are not standalone (unlike PDFs or JPEGs), instead broken up into individual pages. Thus, file carving cannot be applied to DBMS data. 
Carving relational DBMS storage was explored in \cite{stahlberg2007threats, wagner2017database}. More recently, Nissan et al.~\cite{NISSAN2025301929} extended these techniques to the NoSQL domain with ANOC, a tool that automatically carves records and detects tampering directly from database binary files.
However, these carving approaches all perform a ``dead analysis'' on disk images.
Combining the work in this paper with database carving would enable 
a ``live analysis'', such as detecting gaps in database audit logs, similar to malware detection approaches.  

Recent advancements in database memory forensics have focused on validating audit logs by identifying memory access patterns that reflect SQL operations. Wagner et al. \cite{WAGNER2023301567} demonstrated that operations such as full table scans, index accesses, or joins leave distinct repeatable patterns in the DBMS buffer cache and sort area, allowing the identification of query activity that may not appear in the logs due to logging bypasses.

Nissan et al. \cite{nissan2023database} introduced a machine learning-based method for reverse-engineering query activities from memory snapshots, using support vector machines to classify operations like index sort, file sort, or joins based on distinct memory access patterns. Their approach demonstrated high accuracy on MySQL and PostgreSQL DBMS, effectively identifying query types even without persistent logs.

Wagner and Rasin \cite{wagner2020framework} developed a systematic framework to analyze and isolate memory areas across DBMSes, focusing on critical regions such as the I/O buffer, sort area, transaction buffer, and query buffer. Using RAM spectroscopy, they demonstrated how sensitive data, including decrypted information, can persist in these memory areas, highlighting the forensic potential of memory snapshots in analyzing DBMS activity.

\subsection{Network Forensic}
Forensic methods that rely on network traffic analysis face significant challenges in modern environments characterized by encryption and high network traffic \cite{pilli2010network}\cite{dainotti2012issues}. Encryption protocols like TLS protect data in transit but hinder the ability to monitor and reconstruct user activities through network logs \cite{dyer2015marionette}. Even when investigators possess the decryption keys, decrypting large volumes of encrypted traffic is computationally intensive and time-consuming \cite{zhou2022malicious}. The process requires significant computational resources to handle the decryption and reassembly of data streams, especially in high-speed networks where data is transmitted at a rapid rate. Moreover, high-speed networks generate vast amounts of data—over 100 GB daily on a 100 Mbps network—making it impractical for investigators to process and analyze such volumes efficiently \cite{medeiros2020survey}. Handling such massive data sets demands extensive storage capacity and advanced analytical tools, which may not be readily available. The sheer volume also increases the likelihood of missing critical forensic artifacts amidst the vast amounts of irrelevant data \cite{sjostrand2020combatting}.

Network-based forensics also struggles with unreliable timestamps because attackers can manipulate system clocks to obscure their actions, complicating accurate event timeline reconstruction \cite{Atha2023}. Since network analyzers depend on the system time to log events, any alteration of the system clock by an attacker can result in inaccurate or misleading logs, further hindering forensic efforts \cite{scarfone2006nistspecial}. Packet fragmentation further exacerbates the problem; large SQL queries often split across multiple packets, and any loss or reordering of packets makes it difficult to reassemble the full query \cite{pope2013impact}.

\subsection{Database Audit Tools}

Both Peha~\cite{peha1999electronic} and Snodgrass et al.~\cite{snodgrass2004tamper} utilized one-way hash functions to verify audit logs and detect tampering, and Pavlou et al. expanded this work by determining when audit log tampering occurred~\cite{PS08}.  
Rather than detect log tampering, Schneier and Kelsey generated log files impossible to read and impossible to modify by an untrusted user~\cite{schneier1999secure}. Under this framework, an attacker cannot determine if their activity was logged, or which log entries are related to their activity. 
While their mechanisms ensured an accurate audit log with high probability by sending secure hashes to a notarization service, it is ultimately useless if logging has been temporarily suspended by a privileged user. 
MemTraceDB collects database memory artifacts even if their entry in the logs is missing.

An event log can be generated using triggers. However, no DBMS supports \lstinline!SELECT! or SQL DDL (e.g., \lstinline!CREATE! or \lstinline!DROP!)  triggers, making it impossible to log these queries using triggers. The idea of a \lstinline!SELECT! trigger was explored for the purpose of logging~\cite{FR13}. 
MemTraceDB collects query activity found in a DBMS snapshot including both \lstinline!SELECT! and SQL DDL commands.

ManageEngine's EventLog Analyzer~\cite{eventlog} provides audit log reports and alerts for Oracle and SQL Server based on actions, such as user activity, record modification, schema alteration, and read-only queries. However, the Eventlog Analyzer creates these reports based on native DBMS logs. Like other forensic tools, this tool is vulnerable to a privileged user who has the ability to temporarily suspend logs.

Network-based monitoring methods have received significant attention in audit logging research because they provide independence and generality by residing outside of the DBMS. IBM Security Guardium Express Activity Monitor for Databases~\cite{guardium} monitors incoming packets for suspicious activity. If malicious activity is suspected, this tool can block database access for that command or user. 
Liu et al.~\cite{liu2009framework} monitor DBAs and other users with privileged access. Their method identifies and logs network packets containing SQL statements. 

The benefit of monitoring network activity and, therefore, beyond the reach of a DBA, is the level of independence achieved by these tools. On the other hand, relying on network activity ignores local DBMS connections and requires intimate understanding of SQL commands (i.e., an obfuscated command could fool the system).
By contrast, MemTraceDB directly collects evidence of activity that is run against the database instance.

\section{Reliability of Database Logs}
\label{sec:threatmodel}

To establish the forensic necessity for MemTraceDB, I first define the threat model. I assume an attacker who has gained privileged access to the database server, such as a malicious insider or an external threat actor who has escalated their credentials to an administrative level. The primary objective of this attacker is to execute unauthorized commands while erasing any evidence of their actions from standard logging mechanisms. This section details the two primary logging systems the attacker can target to achieve this objective: write-ahead logs (WALs) and audit logs.

\paragraph{Write-Ahead Logs (WALs)}
Write-ahead logs (WALs) record database modifications at a low level to support ACID guarantees. While not designed for auditing, they provide a history of recent table modifications. Although WALs cannot normally be easily modified on a per-record basis and require a special-purpose tool to be read (e.g., PostgreSQL pg\_xlogdump), a privileged user can still manipulate them to hide activity. Some DBMSes allow WALs to be disabled for specific operations, such as bulk loads, allowing an attacker to insert records without leaving a log trace.

Furthermore, most major DBMSes (including Oracle, MySQL, PostgreSQL, and SQL Server) provide administrators with commands to manage the WAL lifecycle. An attacker can exploit this by forcing a switch to a new log file, executing their malicious transactions, and then purging that log file before switching back to the original log stream. In MySQL, where the binary log (binlog) is the equivalent of a WAL, this can be achieved using the following sequence of commands:
\begin{enumerate}
\item \lstinline!FLUSH LOGS;! (Switches from the current log file A to a new log file B)
\item Run malicious SQL operations (These are recorded only in log file B)
\item \lstinline!FLUSH LOGS;! (Switches from log file B to a new log file C)
\item \lstinline!PURGE BINARY LOGS TO 'mysql-bin.log_B';! (Deletes log file B and all evidence within it)
\end{enumerate}

\paragraph{Disabling and Tampering with Audit Logs}
Audit logs are the primary mechanism for recording user activity, but they are entirely under the control of a database administrator. An attacker with privileged access can undermine them in several ways. First, they can temporarily disable audit logging entirely, perform their actions, and then re-enable it, creating a ``blind spot'' in the event history. Second, because audit logs in many systems are often stored as simple, human-readable text files. For example, the PostgreSQL \texttt{pg\_audit} log and the MySQL general query log—an attacker can directly edit these files to surgically remove or alter specific incriminating log entries.

Given that both low-level transaction logs and high-level audit logs can be compromised by a privileged attacker, it is clear that they cannot be trusted as a sole source of forensic evidence. This fundamental unreliability necessitates an out-of-band approach that can reconstruct user activity independently of the database's native logging facilities.
\begin{figure*}[!ht]
\centering
\includegraphics[width=.8\textwidth]{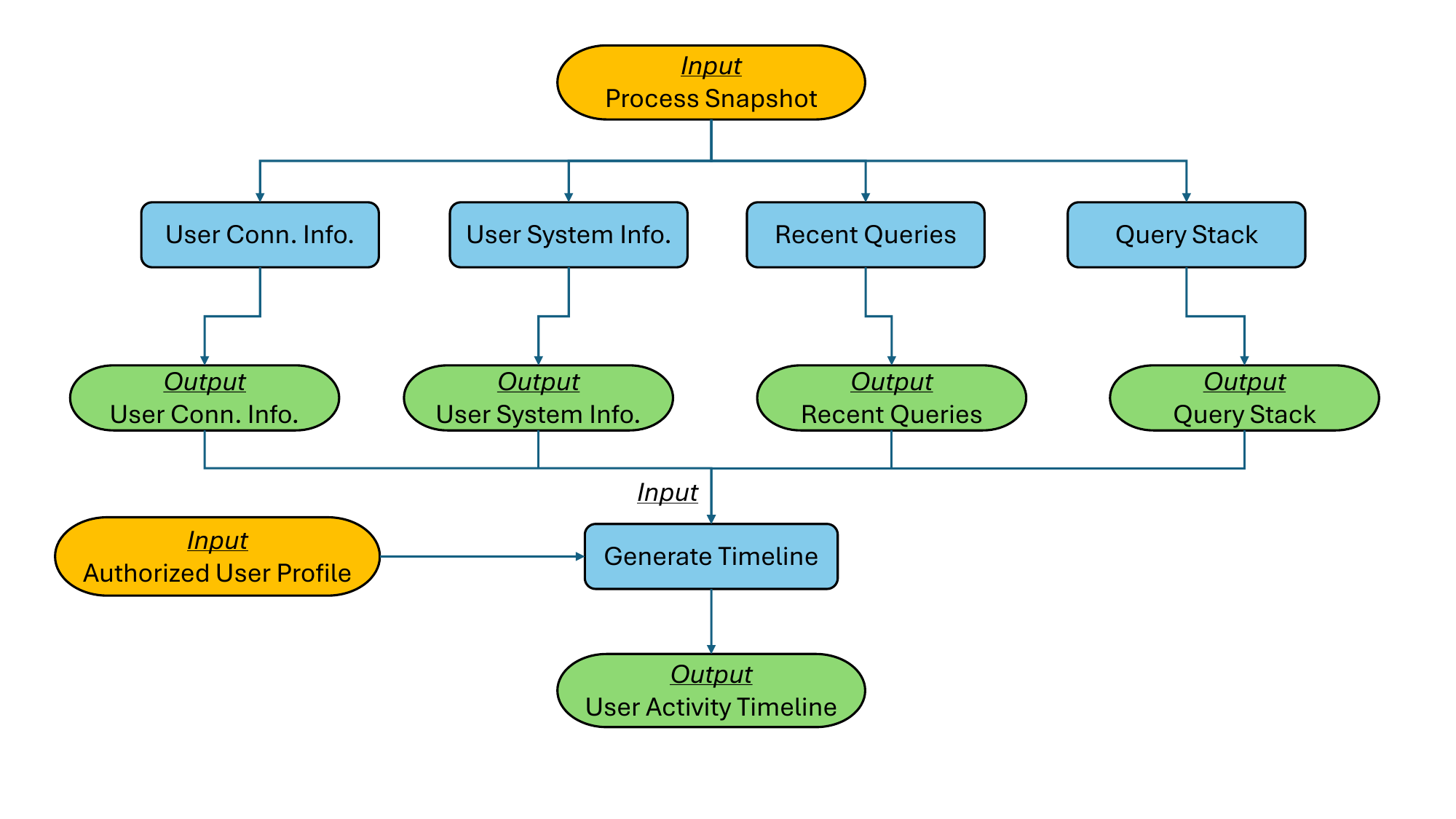}
\caption{The high-level two-stage process of MemTraceDB.}
\label{fig:overview}
\end{figure*}

\section{MemTraceDB Overview}
\label{sec:overview}

Having established the unreliability of conventional logs, I now introduce MemTraceDB, a tool designed to reconstruct a log of user activity by analyzing raw process memory snapshots. The goal is to create a forensically sound timeline from this volatile data. As illustrated in Figure \ref{fig:overview}, MemTraceDB employs a two-stage process to achieve this: \texttt{A) Artifact Extraction} and \texttt{B) Timeline Generation}. This architecture is designed to first isolate discrete pieces of forensic evidence from the noise of the memory snapshot and then to intelligently assemble that evidence into a coherent, chronological narrative.

The first stage, \texttt{Artifact Extraction}, is responsible for carving forensically relevant data structures directly from the process memory snapshot. As input, it takes the raw memory dump and systematically collects key indicators of user activity, including user connection information, user system information, recent queries executed by the user, and the query stack. This process is detailed in Section \ref{sec:artifacts}.

The second stage, \texttt{Timeline Generation}, takes the structured artifacts extracted in the first stage and synthesizes them into a comprehensive user activity log. Using the ActiviTimeTrace algorithm, this component correlates the various pieces of evidence to reconstruct a timeline. The final output is a timeline for each user that details their connection and system information, along with the queries the user \textit{definitely} executed and the queries they \textit{possibly} executed. The procedures for this component are discussed in Section \ref{sec:timeline}.
\begin{figure*}[!ht]
	\centering
	\includegraphics[width=.9\textwidth]{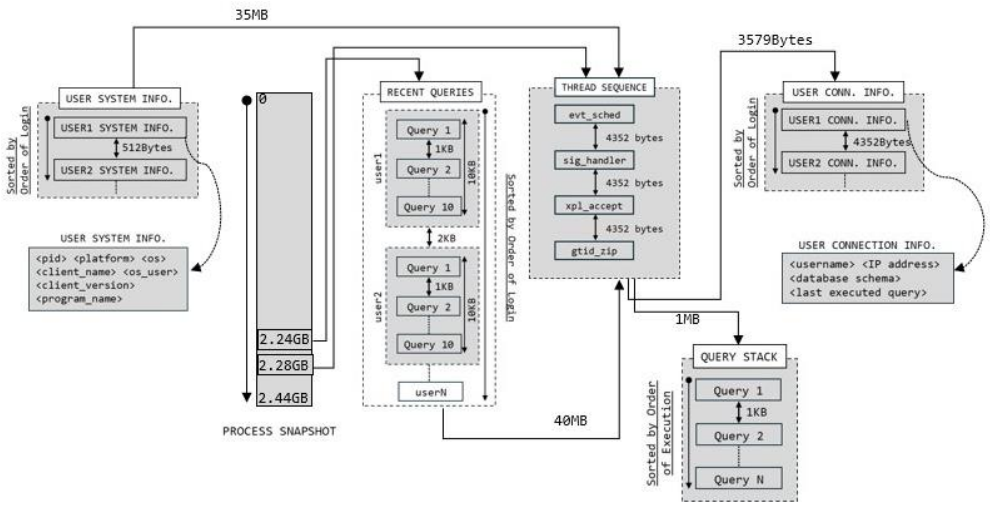}
    \vspace{-4mm}
	\caption{Overview of Artifact Extraction}
    \vspace{-4mm}
    \label{fig:algorithm_outline}
\end{figure*}

\section{Artifact Extraction}
\label{sec:artifacts}

This section explains how MemTraceDB extracts user connection information, user system information, recent queries executed by each user, and the global query stack from MySQL process dumps. Figure \ref{fig:algorithm_outline} provides an overview of this process, and Algorithm \ref{alg:artifacts} describes the procedure step by step. 

\subsection{Initialization}
Line \ref{a1:s}, \textit{S}, is a process snapshot from a MySQL DBMS. Such snapshots can be collected using Procdump v9.0 \cite{procdump} on Windows servers, by copying the relevant process ID file from \textit{/proc/\$pid/mem} on Linux machines, or, in the case of a VM, by taking a full memory snapshot and extracting the relevant process using Volatility \cite{volatility}.  
In the experiments, snapshot sizes typically ranged from 2.1 to 2.6 GB. Although snapshot size may vary depending on workload, this variation does not affect the algorithm. Tests with larger snapshots (e.g., 3 GB and 5 GB) under varying conditions confirmed consistent performance.

Line \ref{a1:t}, \textit{T}, is the thread sequence consisting of the strings \texttt{evt\_sched}, \texttt{sig\_handler}, \texttt{xpl\_accept}, and \texttt{gtid\_zip}. These threads are of particular interest within \textit{S}. Figure \ref{fig:algorithm_outline} illustrates \textit{T}, the `THREAD SEQUENCE' block. MySQL initiates these threads at system startup, and each serves a distinct role:
\begin{itemize} [noitemsep]
    \item \texttt{evt\_sched}: Manages scheduling of database events.
    \item \texttt{sig\_handler}: Processes system-level signals such as interrupts.
    \item \texttt{xpl\_accept}: Handles client connections.
    \item \texttt{gtid\_zip}: Oversees management of Global Transaction Identifiers (GTIDs) used for transaction tracking and replication.
\end{itemize}

Line \ref{a1:block}, \textit{BlockSize\_userconnection}, denotes the distance between one user’s connection block and another, as well as the spacing between individual thread sequence strings. This spacing is consistently 4,352 bytes apart. 

Lines \ref{a1:o1}–\ref{a1:o2} define offsets representing the distances from \textbf{gtid\_zip} to specific data blocks: \textit{Offset\_userconnection} (first user connection block), \textit{Offset\_usersystem} (first user system block), \textit{Offset\_recentqueries} (first recent queries block), and \textit{Offset\_querystack} (query stack block).

Lines \ref{a1:out1}–\ref{a1:out2} initialize empty lists to store results. Specifically, \( C_u \) stores user connection information, \( S_u \) stores user system information, \( Q_u \) stores recent queries for each user, and \( Q_{stack} \) stores all queries found in the global query stack.

\begin{algorithm}[!t]
\caption{Artifact Extraction Algorithm}
\begin{algorithmic}[1]

\State \textbf{1. Initialization}

\State $S \gets$ capture MySQL process snapshot \label{a1:s}
\State $T \gets $ \{\texttt{evt\_sched},\ \texttt{sig\_handler},\ \texttt{xpl\_accept},\ \texttt{gtid\_zip}\} \label{a1:t} 
\State $BlockSize_{userconnection} \gets$ 4352 bytes \label{a1:block}
\State $Offset_{userconnection} \gets $ 3579 bytes \label{a1:o1}
\State $Offset_{usersystem} \gets 35 MB$
\State $Offset_{recentqueries} \gets 40 MB$
\State $Offset_{querystack} \gets 1 MB$ \label{a1:o2}

\State $C_u \gets $ empty list to store user connection information \label{a1:out1}
\State $S_u \gets $ empty list to store user system information
\State $Q_u \gets $ empty list to store recent queries for each user
\State $Q_{stack} \gets $ empty list to store all queries found in the query stack \label{a1:out2}
\State

\State \textbf{2. Identify Thread Sequence} \label{a1:2start}
\State $O_T \gets$ offset of $T \in S$ 
\If{$O_T = \emptyset$} 
    \State
    \Return NULL, NULL, NULL, NULL
\EndIf 
\State $O_{gtid} \gets O_T + 3 \times BlockSize_{userconnection}$ \label{a1:2end}

\State

\State \textbf{3. Collect Connection Information} \label{a1:3start}
\State $O_{C1} \gets O_{gtid} + Offset_{userconnection}$ \label{a1:15} \Comment{Offset of the first user}
\While
\State Extract $u_i, IP_i, DB_i, q_i$
\State $C_u$.append($u_i, IP_i, DB_i, q_i$)
\State $O_{Ci+1} \gets O_{Ci} + BlockSize_{userconnection}$
\EndWhile \label{a1:3end}
\State

\State \textbf{4. Collect User System Information} \label{a1:4start}
\State $O_{S1} \gets O_{gtid} - Offset_{usersystem}$ \Comment{Offset to the first user}
\While{condition?}
\State Extract $PID_i, Platform_i, OS_i, OSUser_i, Client_i, Prog_i$
\State $S_u$.append($PID_i, Platform_i, OS_i, OSUser_i, Client_i, Prog_i$)
\State $O_{Si+1} \gets O_{Si} + 512$
\EndWhile \label{a1:4end}
\State

\State \textbf{5. Collect Recent User Queries} \label{a1:5start}
\State $O_{Q1} \gets O_{gtid} - Offset_{recentqueries}$ \Comment{Offset to the first user query}
\ForEach{user $\in C_u$}     
    \State $User_{queries} \gets$ empty list to store the queries for a user
    \For{\( j = 1 \) to \( 10 \)} 
        \State Extract $Query_{i,j}$
        \State $User_{queries}$.append($Query_{i,j}$)        
        \State $O_{Q_{i,j}} \gets O_{Q_i} + (j - 1) \times 1 KB$
    \EndFor 
    \State $Q_u$.append($User_{queries}$)    
    \State $O_{Q_i} \gets O_{Q1} + (i - 1) \times 12 KB$ 
\EndFor \label{a1:5end}

\State

\State \textbf{6. Collect Query Stack} \label{a1:6start}
\State $O_{Q_{\text{stack}}} \gets O_{gtid} + Offset_{querystack}$
\State \( k \gets 1 \) 
\While{valid query at \( O_{Q_{\text{stack},k}} \)} 
    \State \( O_{Q_{\text{stack},k}} \gets O_{Q_{\text{stack}}} + (k - 1) \times 1 KB \) 
    \State Extract query \( q_k \) at \( O_{Q_{\text{stack},k}} \) 
    \State Add \( q_k \) to \( Q_{\text{stack}} \) 
    \State \( k \gets k + 1 \) 
\EndWhile \label{a1:6end}

\State
\State \textbf{Return} \( C_u, S_u, Q_{u_i}^{(10)}, Q_{\text{stack}} \) \label{a1:41}

\end{algorithmic}
\label{alg:artifacts}
\end{algorithm}

\subsection{Identify Thread Sequence}
Lines \ref{a1:2start}–\ref{a1:2end} initialize the thread sequence location, \( O_T \). If \( O_T \) cannot be found, the results are returned as \texttt{NULL}.  
Figure \ref{fig:algorithm_outline} shows an example with a 2.28 GB offset for \( O_T \).

This sequence serves as a reliable reference point for locating artifacts used to reconstruct user activity. The reason for targeting this sequence is that its components may appear multiple times across the snapshot. However, when they appear in this precise order and with 4,352-byte spacing (\textit{BlockSize\_userconnection}), they provide a stable anchor. Once \( T \) is identified, the offset of \texttt{evt\_sched} defines \( O_T \). From this, the offset for \texttt{gtid\_zip}, \( O_{gtid} \), is calculated as in line \ref{a1:2end}.

\subsection{Collect Connection Information}

Lines \ref{a1:3start}–\ref{a1:3end} describe how connection information for all users is collected; this is illustrated as the `USER CONN. INFO.' block in Figure \ref{fig:algorithm_outline}.
First, $O_{gtid}$ is used as a reference point to locate the connection block for the first user, $O_{C1}$.  
While a valid user connection block is found, the algorithm extracts the username ($u_i$), IP address ($IP_i$), database name ($DB_i$), and last executed query ($q_i$).  
Usernames are identified as alphanumeric strings up to 20 characters, IP addresses through regular expressions such as \texttt{\textbackslash d\{1,3\}(\textbackslash .\textbackslash d\{1,3\})\{3\}} or the keyword \texttt{localhost}, and SQL queries by detecting common SQL command keywords such as \lstinline!SELECT!, \lstinline!INSERT!, or \lstinline!DELETE!.  
The extracted values are appended to $C_u$. The algorithm then advances to the next user connection block using $BlockSize_{userconnection}$.

\subsection{Collect User System Information} 

Lines \ref{a1:4start}–\ref{a1:4end} specify how system information for all users is extracted; this is illustrated as the `USER SYSTEM INFO.' block in Figure \ref{fig:algorithm_outline}.
From $O_{gtid}$, the first user’s system block, $O_{S1}$, is located.  
For each valid block, the algorithm extracts the process ID ($PID_i$), platform ($Platform_i$), operating system ($OS_i$), computer username ($OSUser_i$), client software name ($Client_i$), and the program in use ($Prog_i$).  
These values are found by searching for predefined keys (e.g., \texttt{\_pid}, \texttt{\_platform}, \texttt{\_os}, \texttt{os\_user}, \texttt{\_client\_name}, \texttt{\_client\_version}, \texttt{program\_name}) and recording the associated values in the memory snapshot. Each result is appended to $S_u$. The next system block is then accessed by advancing a fixed, empirically determined offset of 512 bytes.

\subsection{Collect Recent User Queries}
\label{sec:collect_recent_queries}

Lines \ref{a1:5start}–\ref{a1:5end} describe how up to ten most recent queries per user are collected; this is illustrated as the `RECENT QUERIES' block in Figure \ref{fig:algorithm_outline}.
Beginning with $O_{gtid}$, the first query block $O_{Q1}$ is located. For each user in $C_u$, queries are parsed by scanning ASCII strings with the regular expression \texttt{[\textbackslash x20-\textbackslash x7E]\{4,\}}.
Strings matching SQL keywords such as \lstinline!SELECT!, \lstinline!INSERT!, or \lstinline!DELETE! are considered valid and appended to $User_{queries}$. Queries are aligned at 1~KB intervals, which allows stepping to the next entry. The algorithm attempts to collect ten queries per user; if fewer are found before the next user's block begins, it saves all available queries for that user. After the queries are collected, the results are appended to $Q_u$, and the offset then shifts to locate the next user’s block. This offset accounts for up to 10~$\times$~1~KB for queries plus a 2~KB separation, and is scaled accordingly based on the actual number of queries found.

\paragraph{Observations}
When a user executes more than ten queries, the list follows a FIFO replacement policy. Each user’s query block is separated by a 2~KB gap. After finishing the queries for user \( u_i \), the offset for the next block, \( O_{Q_{\text{user}_{i+1}}} \), is calculated as 10~×~1,024 bytes plus the 2~KB separation. This ensures clear separation between users. All queries are consolidated into \( Q_{u_i}^{(10)} \) for later analysis.

\subsection{Extracting the Query Stack}
\label{sec:query_stack}

Lines \ref{a1:6start}–\ref{a1:6end} describe how the global query stack is reconstructed; this is illustrated as the `QUERY STACK' block in Figure \ref{fig:algorithm_outline}.  
Starting from $O_{gtid}$, the first query offset, \( O_{Q_{\text{stack},k}} \), is located. Each query \( q_k \) is parsed using the same method as in Section~\ref{sec:collect_recent_queries}, recorded in \({Q_{\text{stack}}}\), and aligned at 1 KB intervals. The process repeats until the stack is fully extracted.

\subsection{Output}
\label{sec:output}

Finally, Line~\ref{a1:41} returns the four populated lists: $C_{u}$, containing connection details for each user; $S_{u}$, with system information; $Q_{u}$, with recent queries per user; and $Q_{stack}$, the complete query stack across all users. These outputs collectively provide the artifacts necessary to reconstruct user activity from the process snapshot.

\begin{figure*}[!ht]
	\centering
	\includegraphics[width=1\textwidth]{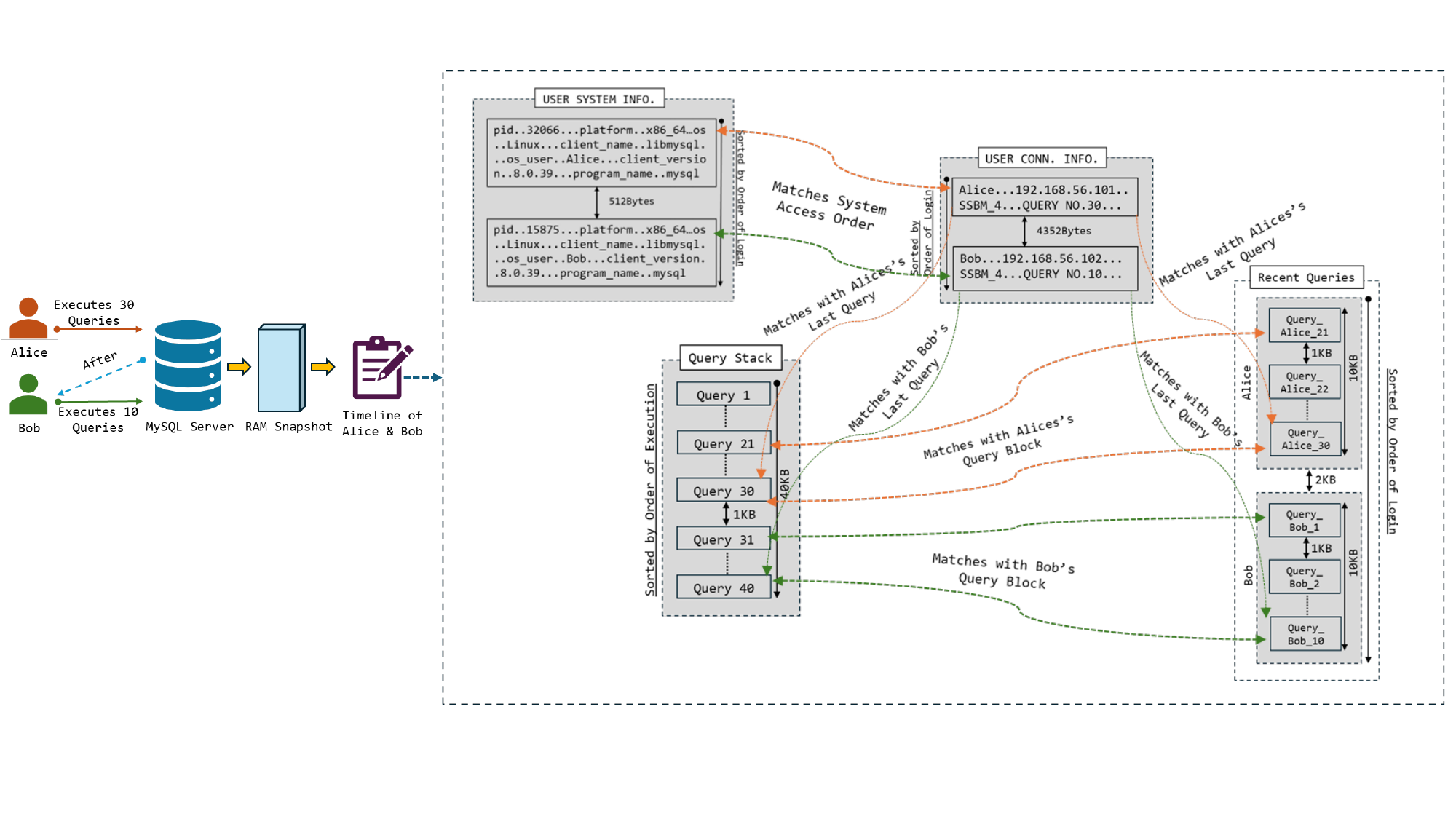}
    \vspace{-2mm}
	\caption{Creating Timeline of Alice \& Bob}
    \vspace{-2mm}
    \label{fig:timeline_ab}
\end{figure*}

\section{Generating a Timeline}
\label{sec:timeline}

This section explains how MemTraceDB reconstructs a timeline of user activity from the output of Algorithm \ref{alg:artifacts}: user connection information (\( C_u \)), user system information (\( S_u \)), the most recent queries executed by each user (\( Q_{u_i}^{(10)} \)), and the global query stack (\( Q_{\text{stack}} \)).  
The result is a timeline of activities for each user, \( \text{Timeline}_i \), showing the order in which queries were executed along with the associated system and connection details. Figure \ref{fig:timeline_ab} provides an overview of this process, and Algorithm \ref{alg:timeline} presents the procedure step by step. 

\begin{algorithm}
\caption{User Activity Timeline Generation Algorithm}
\label{alg:timeline}
\centering
\begin{algorithmic}[1]
\Input \( C_u, S_u, Q_{u_i}^{(10)}, Q_{\text{stack}} \)
\Output \( \{ \text{Timeline}_i \} \) for each user \( u_i \)

\Statex
\Comment{\textbf{Step 1: Verify User's Authenticity}}
\For{each \( u_i \)}
    \State Verify \( \text{login\_seq}_i^{C_u} = \text{login\_seq}_i^{S_u} \)
    \State Check \( (\text{IP}_i, \text{DB}_i) \in L_{\text{conn}_i} \)
    \State Check \( (\text{OS}_i, \text{Platform}_i, \text{Client}_i, \text{Prog}_i) \in L_{\text{sys}_i} \)
    \State \( \text{auth}_i \gets 1 \) \textbf{if} all checks pass, \textbf{else} \( \text{auth}_i \gets 0 \)
\EndFor

\Statex
\Comment{\textbf{Step 2: Identify and Organize User Queries from \( Q_{\text{stack}} \)}}
\For{each user \( u_i \)}
    \State Retrieve the last executed query \( q_i^{(L)} \) from \( C_u \)
    \State Retrieve the 10 most recent queries \( Q_{u_i}^{(10)} \)
    \State Identify the full sequence of queries in \( Q_{\text{stack}} \) using \( q_i^{(L)} \) and \( Q_{u_i}^{(10)} \)
    \State Compute the total number of queries \( n_i \) executed by \( u_i \) based on their positions in \( Q_{\text{stack}} \)
\EndFor

\Statex
\Comment{\textbf{Step 3: Generate the Timeline}}
\For{each user \( u_i \)}
    \State Initialize \( \text{Timeline}_i \gets (S_u, C_u, \{ q_{i,1}, q_{i,2}, \dots, q_{i,n_i} \}) \)
\EndFor

\Statex
\State \textbf{Return} \( \{ \text{Timeline}_i \} \)

\end{algorithmic}
\end{algorithm}

\subsection{Verify User's Authenticity}
In Step 1 of Algorithm \ref{alg:timeline}, the login sequence (\( \text{login\_seq}_i \)) of each user is verified by comparing the login sequences from user system information (\( S_u \)) and connection information (\( C_u \)) to ensure that the system access order matches.  
For example, Figure \ref{fig:timeline_ab} shows that both Alice’s and Bob’s login sequences are consistent between their system and connection details.

The IP address (\( \text{IP}_i \)) and database name (\( \text{DB}_i \)) from \( C_u \) are also checked against the allowed connection list (\( L_{\text{conn}_i} \)) for each user. This ensures that the connection originates from an authorized source and that only permitted databases are accessed.  

Finally, the system information, including the operating system (\( \text{OS}_i \)), platform (\( \text{Platform}_i \)), client software (\( \text{Client}_i \)), and program name (\( \text{Prog}_i \)), is compared with the allowed system configurations list (\( L_{\text{sys}_i} \)). A user is marked as authenticated (\( \text{auth}_i = 1 \)) only if all checks succeed; otherwise, the user is flagged as unauthenticated (\( \text{auth}_i = 0 \)).

\subsection{Identify and Organize User Queries}
In Step 2 of Algorithm \ref{alg:timeline}, queries for each user \( u_i \) are organized using the last executed query \( q_i^{(L)} \) from the connection information \( C_u \) together with the 10 most recent queries \( Q_{u_i}^{(10)} \). These are mapped against the global query stack \( Q_{\text{stack}} \) to reconstruct the user’s activity.  

For instance, as illustrated in Figure \ref{fig:timeline_ab}, Alice executed 30 queries followed by Bob with 10 queries. All 40 queries are stacked in \( Q_{\text{stack}} \) in execution order: Alice’s first 30 queries, then Bob’s 10. From \( Q_{u_i}^{(10)} \) and \( q_i^{(L)} \), queries 21–30 can be identified as Alice’s, while queries 31–40 are attributed to Bob. Because Alice and Bob were both connected during this period, queries 1–20 are inferred to belong to Alice.  

By matching \( q_i^{(L)} \) and \( Q_{u_i}^{(10)} \) within \( Q_{\text{stack}} \), the complete sequence of queries executed by each user is recovered. The total number of queries \( n_i \) executed by each user is then computed from their positions in \( Q_{\text{stack}} \).

\subsection{Generate the Timeline}
In Step 3 of Algorithm \ref{alg:timeline}, the activity timeline (\( \text{Timeline}_i \)) for each user is constructed by combining system information (\( S_u \)), connection information (\( C_u \)), and the complete set of executed queries \(\{ q_{i,1}, q_{i,2}, \dots, q_{i,n_i} \}\).  

Each resulting tuple \( (S_u, C_u, \{ q_{i,1}, q_{i,2}, \dots, q_{i,n_i} \}) \) provides a structured record of user activity, including platform, operating system, client version, program, IP address, database accessed, and the sequence of executed queries.  

For example, if Alice executed 30 queries, her timeline would contain \(\{ q_{i,1}, \dots, q_{i,30} \}\) combined with her system and connection details. The final output is the set of timelines \(\{ \text{Timeline}_i \}\), one for each user, each representing both the execution sequence and the environment in which those actions occurred.

\section{Experiments}
\label{sec:experiments}
\begin{table}[!t]
\centering
\small
\begin{tabular}{|l|c|}
\hline
\textbf{Table} & \textbf{Records} \\ \hline

Date     & 2556    \\ \hline
Supplier  & 20K     \\ \hline 
Customer  & 300K    \\ \hline 
Part      & 800K    \\ \hline 
Lineorder & 60M     \\ \hline
Total     & 61M     \\ \hline
\end{tabular}
\caption{SSBM Scale 10 Table Sizes}
\vspace{-8mm}
\label{table:ssbm_table_scale10}
\end{table}

\begin{table*}[!htb]
\centering

\begin{tabularx}{\textwidth}{| l | >{\raggedright\arraybackslash}X | >{\raggedright\arraybackslash}X |} 
\hline
\rowcolor{gray!30}\textbf{Operation} & \textbf{Summary} & \textbf{SQL Template} \\ 
\hline

DDL Commands & Used to define, modify, and manage the structure of a database. & 
\lstinline!CREATE [table/view cond.]! \newline \lstinline!DROP [table_name/view_name]! \\ 
\hline

Full Table Scan & Scan the entire table to retrieve a record(s), without utilizing indexing or optimization techniques. & 
\lstinline!SELECT * FROM [table_name]! \\ 
\hline

Index Sort & A record(s) is obtained by using a (often) B-Tree index to identify a pointer(s) that links to the record(s). & 
\lstinline!SELECT * FROM [table_name]! \newline \lstinline!ORDER BY [indexed_column]! \\ 
\hline

File Sort & It is used when a sorting operation can't utilize index access. & 
\lstinline!SELECT * FROM [table_name]! \newline \lstinline!ORDER BY [non_indexed]! \\ 
\hline

Join & Join operation can be hash join, two nested for-loops or merge join. & 
\texttt{SELECT [table\_x \& table\_y]} \newline 
\lstinline!FROM [table_x] JOIN [table_y]! \newline 
\lstinline!ON table_x.ID = table_y.ID! \\ 
\hline

Filter & Filter rows according to the criteria specified in the \lstinline!WHERE! clause condition. & 
\lstinline!SELECT * FROM [table_name]! \newline \lstinline!WHERE [where_cond.]! \\ 
\hline

Aggregate & Commonly used with \lstinline!GROUP BY! clause to group values into subsets. & 
\lstinline!SELECT [column],! \newline
\lstinline![aggregate_cond.]! \newline 
\lstinline!FROM [table_name]!
\lstinline!GROUP BY [column]! \\ 
\hline
\end{tabularx}
\caption{Query Workload}
\label{tbl:workload1}
\end{table*}

\paragraph{Purpose}
The purpose of the experiments is to determine the frequency of taking process snapshots to build user activity timeline.
\paragraph{Procedure}
\label{sec:user_sim_setup}
I used two different procedures to simulate user activities on a DBMS: (1) multiple users on different virtual machines (VMs) and (2) multiple users on the same virtual machine (VM). These two approaches were chosen to ensure consistent results while optimizing system resources. Simulating user interactions on the same system provided comparable outcomes to simulating them on different computers connected to the server over the network, with the added benefit of reducing the number of VMs required. This approach minimized resource usage and maintained system stability, ensuring efficient testing. The only difference was in the IP addresses, where, instead of distinct IPs like 192.168.1.1 and 192.168.1.2, the IP Address appeared as \texttt{localhost} when simulating multiple users on the same machine.

\subsection{Experimental Setup}
\paragraph{Dataset}
The experiments were conducted using Scale 10 of the Star Schema Benchmark (SSBM) \cite{o2009star}, as outlined in Table \ref{table:ssbm_table_scale10}. The SSBM simulates a data warehouse environment, providing realistic data distributions through a synthetic data generator that produces datasets at different scale levels.

\paragraph{Workload} I generated SQL queries to evaluate the performance of MemTraceDB using a Python script that followed the query workload template from Table \ref{tbl:workload1} and utilized the SSBM scale 10 dataset (Table \ref{table:ssbm_table_scale10}). The generated SQL queries were then saved in a \texttt{Query.sql} file.

\paragraph{User Interaction Simulation (Expect)} I used an Expect script to simulate user interactions. Expect is a scripting tool that automates interactions with command-line programs \cite{expect}, enabling me to simulate SQL queries being executed by different users as if they were interacting with the MySQL server in real time. To connect from Alice and Bob's VMs to the MySQL VM remotely, I used the command template: \texttt{mysql -h \$host -u \$user -p}, while for local connections, I omitted the \texttt{-h \$host} option from the command. 
   
\paragraph{User Interaction Simulation (Bash)} A Bash script was used to launch the Expect script for both single and multiple users, simulating various user activity scenarios. The script read SQL queries from a file, \texttt{Query.sql} and assigned them to each user (e.g., \texttt{user1}, \texttt{user2}). It initiated query execution in parallel, with a 15-second delay between queries, mimicking the behavior of multiple legitimate users interacting with the database server simultaneously.

\subsection{Exp. 1: Network Users}
\label{sec:exp1_network_users}

\paragraph{Purpose}
The purpose of Exp 1 is to demonstrate that generating a user activity timeline is not different whether the user is connected to the server over the network or directly via localhost.

\subsubsection{Setup}
\label{sec:multi_vm_setup}
I created three separate virtual machines (VMs) using VirtualBox \cite{virtualbox}: two for the users Alice and Bob, and one dedicated to hosting the MySQL server. Each VM runs Ubuntu 20.04 LTS, with MySQL version 8.0.39 installed. I allocated 8 GB of RAM (total system RAM 64 GB) and 4 processors (AMD Ryzen 7950X3D) for each VM. To simulate a real-world scenario where Alice and Bob interact with a remote MySQL server, I configured the network settings of each VM to behave as separate systems. The network adapter for each VM was set to \texttt{Bridged Adapter} mode in VirtualBox, allowing the VMs to obtain unique IP addresses from the local network, making them accessible to one another as if they were on different physical machines.

\subsubsection{Procedure} 
\label{sec:multi_vm_procedure}

I conducted the experiment in two steps, first a single-user experiment followed by a two-user experiment. Before each experiment, whether with a single user (Alice) or two users (Alice and Bob), I first defined the usernames, passwords, and, for remote VMs, the IP address of the MySQL server. I used the workload template (as shown in Table \ref{tbl:workload}) to generate a total of 18,000 SQL queries and utilized these generated queries for different experimental scenarios, adjusting the number of queries according to each scenario. To simulate user interactions, I employed Bash and Expect scripts, as explained in Section \ref{sec:user_sim_setup}. After each experiment, I used ProcDump to take a process snapshot and used it as input for MemTraceDB. MemTraceDB generated a user activity timeline (Timeline.txt) along with additional output files, including User\_Connection\_Info.txt, User\_System\_Info.txt, Recent\_Queries.txt, and Query\_Stack.txt. These files provide detailed outputs for each block, as explained in Section \ref{sec:artifacts}, which are used by MemTraceDB to reconstruct the timeline.

The following outlines the individual experimental scenarios. I created these scenarios to test the timeline generation of MemTraceDB in different conditions. Before each experimental scenario, i.e., S1, S2, S3, and S4, I cleared the user's query cache using \texttt{FLUSH USER RESOURCES} to ensure there was no query history.  However, the query cache was not cleared before steps within each scenario, i.e., S3.1, S4.1, S4.2, and S4.3,  so the previous session's queries remained cached.

\begin{enumerate}[leftmargin=25pt, itemsep=0.5em, align=left]
    \item[\textbf{S1 (Single User):}] Alice executed 18,000 queries.
    \vspace{-2mm}

    \item[\textbf{S2 (Two Users):}] Alice \& Bob each executed 9,000 queries, totaling 18,000 in parallel.
    \vspace{-2mm}

    \item[\textbf{S3 (Two Users):}] Alice executed 9,950 queries.
    \vspace{-2mm}
    \begin{enumerate}[left=0pt, labelwidth=!, itemsep=0.5em]
        \item[\textbf{S3.1:}] Bob logged in and executed 100 queries.
    \end{enumerate}
    \vspace{-2mm}

    \item[\textbf{S4 (Two Users):}] Alice executed 1000 queries.
    \vspace{-2mm}
    \begin{enumerate}[left=0pt, labelwidth=!, itemsep=0.5em]
        \item[\textbf{S4.1:}]Alice logged out. After that, Bob logged in and executed 1000 queries.
        \vspace{-2mm}
        \item[\textbf{S4.2:}] Bob logged out. After that, Alice logged in again and executed 1000 queries.
        \vspace{-2mm}
        \item[\textbf{S4.3:}] Alice logged out. After that, Bob logged in again and executed 1000 queries.
    \end{enumerate}
    \vspace{-2mm}
\end{enumerate}

\paragraph{Single User}
For the single-user experiment (i.e., \textbf{S1}), I used Alice's system and started with 1,000 queries, gradually increasing the number to 18,000 unique queries to observe the system's behavior and assess its capacity to hold the necessary information for generating a user activity timeline using MemTraceDB.  I took a process snapshot after every 1,000 queries.  

\paragraph{Two Users}
For the two-user experiment where both users execute queries in parallel (i.e., \textbf{S2}), I divided the 18,000 total queries equally, assigning 9,000 queries to each user. I started with 1,000 queries per user and gradually increased the number to 9,000 unique queries per user to observe the system's behavior and assess its capacity to hold the necessary information for generating a user activity timeline using MemTraceDB. 
After every 1,000 queries per user (i.e., 2000 total), I took a process snapshot. To find out any user's large number of query execution effect on the query recovery (i.e., \textbf{S3}), I started with Alice and executed 9,950 queries. After that, I started with Bob and executed 100 queries (i.e., \textbf{S3.1}).

To simulate a scenario where a user logs in, executes queries, and then logs out, I first logged into Alice's session and ran 1,000 queries (i.e., \textbf{S4}). Next, to simulate another user following the same steps (i.e., \textbf{S4.1}), I logged out of Alice's session, then logged into Bob's session and executed 1,000 queries. I then repeated this process for a returning user (i.e., \textbf{S4.2}) by logging out of Bob's session and logging into Alice's session again, executing 1,000 queries. Finally, I performed the same simulation for Bob (i.e., \textbf{S4.3}) as in S4.2.

\subsubsection{Result \& Discussion}
\label{sec:result_discussion_remote}

\if false
\begin{table}[h!]
\centering
\begin{tabular}{|c|c|c|c|c|}
\hline
\multirow{2}{*}{\textbf{Exp.}} & \multicolumn{2}{c|}{\textbf{Query Execution}} & \multicolumn{2}{c|}{\textbf{Query Map}} \\
\cline{2-5}
 & \textbf{Alice} & \textbf{Bob} & \textbf{Alice} & \textbf{Bob} \\
\hline
S1 & 18,000 & - & 10,007 & - \\
\hline
\hspace{5mm} S1.1 & 18,100 & - & 10,007 & - \\
\hline
S2 & 9,000 & 9,000 & 11 & 11 \\
\hline
\hspace{5mm} S2.1 & - & 100 & 11 & 11 \\
\hline
\hspace{5mm} S2.2 & 100 & - & 11 & 11 \\
\hline
S3 & 9,950 & - & 9,950 & - \\
\hline
\hspace{5mm} S3.1 & - & 100 & 9,949 & 56 \\
\hline
\end{tabular}
\caption{Result Summary}
\label{tbl:2_users_summary}
\end{table}
\fi

\begin{table*}[!ht]
	\centering
	\includegraphics[width=1\textwidth]{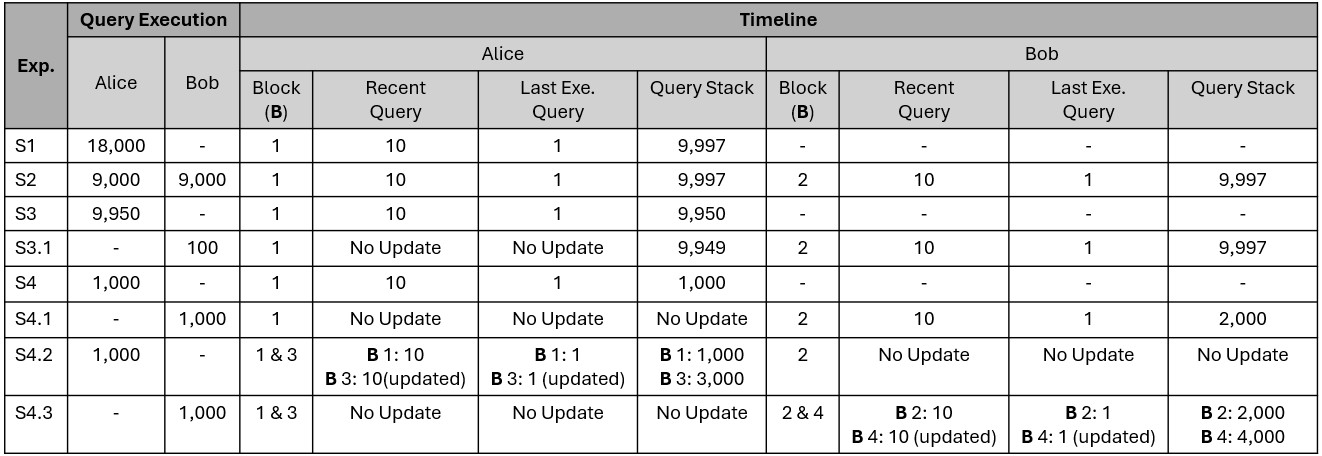}
        \vspace{-2mm}
	\caption{Summary of Exp 1}
        \vspace{-2mm}
    \label{tbl:summary_exp_1}
\end{table*}

\paragraph{Single User}

Table \ref{tbl:summary_exp_1} summarizes the results of query mapping for building the user activity timeline.\\

\textbf{S1.} After at around 9997 queries execution, MySQL began replacing the first query in the query stack with the latest one, and approximately 9,996 previous queries remained unchanged. I also observed that at around query number 16,240, MySQL stopped updating both the most recent queries and the last executed query. High memory usage was also detected, and MySQL began utilizing the SWAP file. This could have contributed to the failure to update queries, as MySQL may have started offloading some of its work to the SWAP file. MemTraceDB generated the timeline of Alice by creating \texttt{Block 1} and mapped the user system information, user connection information, last executed query, and the most recent queries around query number 16,240. It mapped 9,997 queries to query stack, where 9,996 queries remained unchanged, with only the first query being replaced by the latest one, as reflected in the process snapshot.

\paragraph{Two Users}

Table \ref{tbl:summary_exp_1} summarizes the results of query mapping for building the user activity timeline.\\ 

\textbf{S2.} When two users execute queries in parallel, the findings were consistent with the single-user experiment: after the total number of queries exceeded 9,997, MySQL began replacing the first query in the stack with the most recent one, regardless of which user executed it. Between the two users, I observed that MySQL stopped updating Alice's last executed query and recent queries block at approximately 7,593 queries, while Bob’s stopped at around 7,909 queries. Beyond the 9,997-query limit, MemTraceDB could not separate the query stack based on individual users. This occurred because the last executed queries for both Alice and Bob were no longer present in the query stack, making it impossible to identify how many queries a user may have executed from the query stack. Once MySQL reached the 9,997-query limit, it started replacing the oldest query with the latest query executed by either user, while the remaining 9,996 queries stayed the same.  MemTraceDB created \texttt{Block 1} for Alice and \texttt{Block 2} for Bob in the timeline and mapped their respective recent queries and last executed query as found in the recent query block and user connection information block. It assigned all 9,997 queries to both Alice and Bob's Query Stack section in the timeline.\\

\textbf{S3.} When Alice executed the majority of the queries (i.e., 9,950), MemTraceDB was able to map all of her queries in the timeline.\\

\textbf{S3.1.} When Bob logged in and executed 100 queries, similar to previous experiments, the combined queries exceeded 9,997 and began overwriting entries at the location of Alice's first query.  In the timeline, I found a total of two blocks: \texttt{Block 1} for Alice and \texttt{Block 2} for Bob. Since Bob replaced one of Alice's queries, I observed that Alice's query stack now holds 9,949 queries. For Bob, MemTraceDB mapped 9,997 queries to his query stack. By examining the query stack, I can infer the number of queries Alice may have executed, identify when she stopped executing queries, and confirm that all queries following Alice's query stack entries were executed by Bob.\\

\textbf{S4.}  MemTraceDB generated timeline by creating \texttt{Block 1} and mapping the queries to their respective sections: 10 recent queries, 1 last executed query, and 1,000 queries in the query stack.\\

\textbf{S4.1} MemTraceDB generated timeline by creating a block for Bob, \texttt{Block 2}. It mapped Bob's recent queries and last executed query to \texttt{Block 2}. However, it mapped 2,000 queries to \texttt{Block 2's} query stack, with the first 1,000 queries originating from Alice. By observing \texttt{Blocks 1} and \texttt{2} in the timeline, it remains possible to identify which queries were not executed by Alice and which were definitely executed by Bob, as the timeline correctly separated Alice's query stack in \texttt{Block 1}. Additionally, as Alice's system information were missing from the Block 1 of timeline, I can also identify that, at the time of taking process snapshot, Alice had already logged out, and Bob was the only active user.\\

\begin{figure}[!htb]
        \centering
	\includegraphics[width=0.2\textwidth]{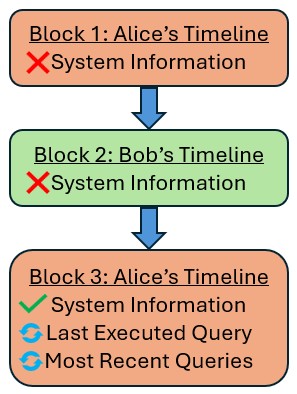}
    \caption{Alice \& Bob's Blocks in the Timeline}
	\label{fig:alice_bob_blck123}
\end{figure}

\textbf{S4.2} From this experiment, I identified a new block in the timeline, \texttt{Block 3}, for Alice. MemTraceDB mapped Alice's updated recent queries and last executed query to \texttt{Block 3}. However, it mapped a total of 3,000 queries to Alice's query stack: the first 1,000 from Alice's \texttt{Block 1} and the next 1,000 from Bob's \texttt{Block 2}. In the timeline, I found three blocks in total: \texttt{Block 1} for Alice, \texttt{Block 2} for Bob, and \texttt{Block 3} for Alice. System information was missing in \texttt{Blocks 1} and \texttt{2}, while only \texttt{Block 3} contained Alice's system information, along with the updated recent query and last executed query, as shown in Figure \ref{fig:alice_bob_blck123}. From this timeline, I can deduce that Alice logged in first, executed 1,000 queries, and then logged out. Bob then logged in and may have executed up to 2,000 queries before logging out. Finally, I can conclude that Alice logged in again, and her connection was active at the time of taking process snapshot, potentially executing up to 3,000 queries while she was logged in.\\

\textbf{S4.3} The findings from this experiment are similar to \textbf{S4.2}. I observed a new block, \texttt{Block 4}, for Bob. MemTraceDB mapped Bob's updated recent queries and last executed query to \texttt{Block 4} and mapped a total of 4,000 queries to Bob's query stack: the first 1,000 from Alice's \texttt{Block 1}, the second 1,000 from Bob's \texttt{Block 2}, and the third 1,000 from Alice's subsequent session. Based on \textbf{S4.3}, I can identify that Alice logged in first, executed 1,000 queries, and then logged out. Bob then logged in and may have executed up to 2,000 queries before logging out. Alice logged in again and may have executed up to 3,000 queries before logging out. Finally, Bob logged in again, and his connection was active at the time of taking process snapshot, potentially executing up to 4,000 queries while he was logged in.

\subsection{Exp. 2: Local Users}
\label{sec:exp2_local_users}

\paragraph{Purpose}
The purpose of Exp 2 is to analyze the effect of increasing the users on query recovery.

\subsubsection{Setup}
To simulate multiple users accessing the MySQL server on the same VM that also hosts the MySQL server, I created 100 unique MySQL users. A Python script was used to connect to the MySQL server and iteratively generate users (\texttt{user1} to \texttt{user100}). I used the same MySQL server VM as described in Section \ref{sec:multi_vm_setup}. As all users are on same VM, here, instead of using distinct IP addresses, I used \texttt{localhost} for all users to connect to the server.

\subsubsection{Procedure}
I followed a similar procedure as outlined in Section \ref{sec:multi_vm_procedure}. The only difference in this case was that, instead of each user having a distinct IP address, all users shared the same local system, using \texttt{localhost} as the connection point. 
Based on the memory query threshold I found in Experiment 1,
I used the following formula to evenly distribute the queries among users, ensuring that the total number of queries from all users did not exceed the limit of 9995:

\begin{equation}
Q_n = \frac{9995}{n}
\label{eq:queries_per_user}
\end{equation}

Here, \(Q_n\) is the number of queries per user, and \(n\) is the number of users.

For the different experimental scenarios, the same procedures were applied as in the previous section \ref{sec:exp1_network_users}. Before each experimental scenario (i.e., S5, S6, and S7), I cleared the user's query cache using \texttt{FLUSH USER RESOURCES} to ensure no residual query history.

\begin{enumerate}[leftmargin=25pt, itemsep=0.5em, align=left]
    \item[\textbf{S5 (Multiple Users):}] Logged in and executed queries sequentially from user1 to user30, allocating queries to each user according to Formula \ref{eq:queries_per_user}.
    \vspace{-2mm}
    \item[\textbf{S6 (Multiple Users):}] 30 users logged in, each executing 10 queries in parallel.
    \vspace{-2mm}
    \item[\textbf{S7 (Multiple Users):}] 100 users, each executing 10 queries, in parallel.
\end{enumerate}

\subsubsection{Result \& Discussion}
\label{sec:result_discussion_local}

\begin{table*}[!ht]
	\centering
	\includegraphics[width=.9\textwidth]{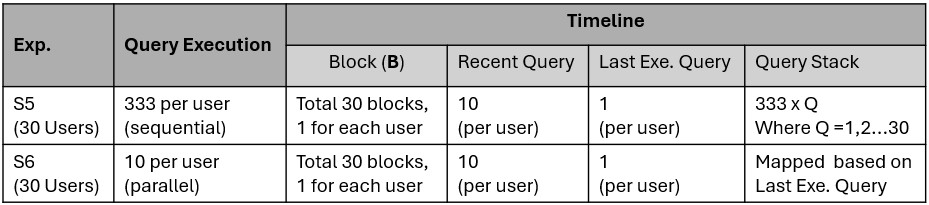}
        \vspace{-2mm}
	\caption{Summary of Exp 2}
        \vspace{-2mm}
    \label{tbl:summary_exp_2}
\end{table*}

\textbf{S5.} The experimental results were similar to Exp 1. MemTraceDB created 30 blocks in the timeline, assigning each block to a user in the order of their login. It mapped 10 recent queries and 1 last executed query to each user's block. As each user logged in and executed 333 queries sequentially, the queries accumulated in the stack as the first 333 queries from user1, the next 333 from user2, and so on. I observed that in each user's query stack on the timeline, MemTraceDB mapped 333 queries for user1, 666 for user2, continuing this pattern up to the 9,990th query for user30. From the timeline, it is possible to identify the total number of queries each user may have executed.\\

\textbf{S6.} From this experiment, it was possible to identify all the queries each user executed. MemTraceDB created 30 blocks in the timeline, assigning each block to a user in the order of their login. It mapped 10 recent queries and 1 last executed query to each user's block. Since the queries were executed in parallel, MemTraceDB assigned a varying number of queries in each query stack based on the last executed query. Nevertheless, it remains possible to determine the number of queries each user executed and their order of execution. With a total of 300 queries across all users' query stacks and 10 recent queries per user (totaling 300 recent queries), it is possible to accurately identify both the number and order of queries executed by each user.\\

\textbf{S7.} In my final experiment, I tested the system with 100 users, assigning 10 queries to each user. The users logged in sequentially, from \texttt{user1} to \texttt{user100}. After a few queries, the system slowed down drastically. To prevent a system crash, I shut down the MySQL clients for each user and took a process snapshot, which resulted in a snapshot size of around 9.3 GB. This size was unusual, considering my system had only 8 GB of RAM. MemTraceDB could not find the necessary anchor points to extract the information, leading me to believe that the process snapshot structure was damaged.

\subsection{Snapshot Frequency Based on Query Limits}

To ensure that the total number of executed queries does not exceed the system's query limit of 9,997, I propose Formula \ref{eq:snapshot_formula} for determining the optimal time to take process snapshots.

I assume that each user can execute a maximum of 180 queries per hour, which equates to 3 queries per minute. This limit can be enforced in MySQL by using the \texttt{MAX\_QUERIES\_PER\_HOUR} option when defining user privileges:

\begin{verbatim}
GRANT USAGE ON *.* TO 'username'@'host' WITH
MAX_QUERIES_PER_HOUR 180;
\end{verbatim}

This limits the number of queries a user can execute to 180 per hour, thereby ensuring that no user exceeds the predefined query rate of 3 queries per minute.

Let:
\begin{itemize}
    \item \(n\) be the number of active users.
    \item \(q_{\text{user}}\) be the query limit per user, set at 3 queries per minute (or 180 queries per hour).
    \item \(q_{\text{total}} = 9997\) be the total query limit before a process snapshot is required.
\end{itemize}

The rate at which queries are executed by all users is given by:
\[
q_{\text{rate}} = 3 \times n \quad \text{queries/minute}
\]
where each user executes up to 3 queries per minute.

The total time until a snapshot should be taken is calculated by dividing the total query limit \(q_{\text{total}}\) by the query execution rate \(q_{\text{rate}}\), resulting in the formula:

\begin{equation}
t_{\text{snapshot}} = \frac{9997}{3 \times n}
\label{eq:snapshot_formula}
\end{equation}

where \(t_{\text{snapshot}}\) is the time (in minutes) between process snapshots.

For example, if there are 10 active users, the query rate is:
\[
q_{\text{rate}} = 3 \times 10 = 30 \quad \text{queries/minute}
\]
Substituting into the formula:
\[
t_{\text{snapshot}} = \frac{9997}{30} \approx 333 \quad \text{minutes}
\]
Thus, for 10 users, the system should take a process snapshot approximately every 333 minutes (about 5.5 hours) to ensure that the total number of executed queries does not exceed the system's limit of 9,997.

\subsection{Observation}
During my experiments, I found that the only clean SQL queries not affected by previous query artifacts were located in the query stack block. If the initial query's text size was significantly longer than the subsequent queries, the queries in the most recent queries block and the last executed query could become heavily corrupted. MemTraceDB occasionally failed to correct highly corrupted last executed query if they did not closely match any query in the query stack, which MemTraceDB uses to repair damaged queries.

I also observed that if a user executed 6 or fewer queries, additional commands were found in both the most recent query block and the query stack, even though the user had not explicitly run them. These were system-level commands, such as \texttt{SELECT @@version\_comment LIMIT 1}, \texttt{SELECT DATABASE()}, \texttt{SHOW DATABASES}, and \texttt{SHOW TABLES}. Since these commands were initiated by the system rather than the user, MemTraceDB filtered them out. Additionally, I discovered that when connecting to the database server using the Python script, the system information of the connecting user did not appear. System information was only captured when connecting via the MySQL client.

As mentioned in Sections \ref{sec:result_discussion_remote} \& \ref{sec:result_discussion_local}, even after closing the user connections, I continued to find the previous 9,996 queries in the query stack that were not updating in the memory snapshot. This persistence is likely due to several factors, including the behavior of the operating system and MySQL's internal memory management. Operating systems like Linux do not always immediately clear memory after a process finishes using it, instead retaining the memory allocated to a process until it needs to be reused. This explains why query data may still be present in memory after closing the connection. Additionally, MySQL uses caching and buffering techniques, such as the InnoDB buffer pool, which stores frequently accessed query data in memory.

\subsection{Summary of Experimental Findings}
\label{sec:summary_findings}

The experiments in this section reveal several key characteristics and limitations of using memory analysis for timeline reconstruction in MySQL. Three primary findings stand out. First, the MySQL query stack has a finite operational capacity of approximately 9,997 queries. Once this threshold is exceeded, the system begins to overwrite the oldest query with the newest one, while the rest of the stack remains largely static.

Second, this finite capacity directly impacts query attribution in multi-user scenarios. While MemTraceDB can successfully separate user activities under moderate loads, the attribution becomes ambiguous when the total query count surpasses the stack's limit or when a high number of users are executing queries in parallel.

Finally, the experiments demonstrate that artifacts from logged-out user sessions persist in memory. These sessions can be identified forensically by the presence of a user's connection block and query history but the corresponding absence of their active system information block. However, the experiments also highlighted a scalability threshold; the system became unstable with 40 or more simultaneous users, leading to corrupted process snapshots and preventing successful artifact extraction. These findings inform the practical application of MemTraceDB and the necessary frequency of snapshot collection outlined previously.

\section{Comparison of Memory Analysis and Network Packet Analysis}
In forensic investigations, determining user activity and detecting suspicious queries often requires monitoring data flow. Traditionally, network packet analysis has been the primary method for this purpose. However, with the increasing use of encryption protocols and the complexity of modern network environments, memory analysis is emerging as a more efficient alternative. This section compares the two approaches and highlights the advantages of using memory analysis, particularly with MemTraceDB, over network packet analysis.

\subsection{Challenges in Query Analysis and Encryption Overhead}

In practice, network packet analysis involves several challenges, particularly when queries are sent over encrypted channels, such as those using TLS 1.2 or TLS 1.3 in MySQL. These queries require decrypting the entire communication stream, which includes both queries and their results. The decryption process adds significant overhead due to the computational complexity of encryption algorithms like AES-256-GCM and RSA-2048 \cite{coarfa2006performance, clark2013sok}. For instance, decrypting MySQL queries with SSL/TLS can increase query times by 34\% to 36\% in typical configurations \cite{planetmysql2013l}. This overhead is further compounded in high-traffic environments or when multiple encryption sessions are in use, requiring more processing power.

Moreover, packet fragmentation adds complexity to query analysis. Database queries are often broken into multiple packets, which need to be reassembled before they can be decrypted and analyzed \cite{kurose2022computer}. Since network captures typically do not separate query and response packets, isolating and analyzing only the queries becomes a difficult task, adding to the computational overhead.

\subsection{Protocol Complexity and Packet Fragmentation}

In addition to encryption overhead, the complexity of communication protocols like MySQL further complicates packet analysis. Queries and responses are often intertwined within the same packet stream, requiring a deep understanding of the protocol to properly interpret the data \cite{mysql_protocol}. Furthermore, the TCP/IP protocol introduces its own overhead through packet headers, acknowledgments, and error-checking mechanisms, which add extra layers of data that need to be processed \cite{rfc793}. This increases the time and complexity involved in isolating and analyzing the queries from the overall network traffic.

\subsection{Memory Analysis with MemTraceDB}

Memory analysis, in contrast, bypasses many of these issues. Since memory captures data in its complete form, it eliminates the need for traffic decryption, packet reassembly, and protocol handling. A single MySQL process snapshot averages around 2.44 GB and can be generated in approximately 20 seconds. \textbf{MemTraceDB} can then analyze the snapshot and generate a complete user activity timeline in an average of \textbf{26 seconds}.

\paragraph{No Decryption Required} Since memory snapshots capture data in its decrypted form, there is no need for additional decryption during analysis. This eliminates decryption overhead, significantly speeding up the process and making it more computationally efficient.

\paragraph{No Packet Reassembly} Memory captures the entire query as a complete unit, removing the need for reassembling fragmented packets, which is typically required in network packet analysis. This reduces the time and complexity associated with handling fragmented network traffic.

\paragraph{No Protocol Overhead} Memory snapshots are free from network protocol headers and transmission issues, making the queries easier to extract and analyze. This removes the need to filter out irrelevant network protocol data, further streamlining the analysis process.

\subsection{Challenges of Memory Analysis}

While memory analysis provides a direct, decrypted view of data, it has limitations that can impact forensic accuracy. Data in memory is highly transient, meaning it can be lost if the system experiences a power outage or restarts unexpectedly. This volatility requires timely acquisition, as any delay risks losing critical information. When analyzing a MySQL process snapshot, there is complexity in handling user-specific query data accurately. While I can estimate the approximate number of queries each user may have executed, I found that accurately mapping all queries to individual users proved impossible in most experiments. In many cases, it was not feasible to determine the exact starting point of a user’s query execution, as queries often get overwritten in the recent query block, and the query stack lacks indicators to trace where a user’s queries began. Additionally, to avoid exceeding the 9,997-query threshold, I had to take process snapshots and periodically clear the query cache. Scalability is another issue, as performance degrades with large numbers of users. In my experiment with 100 users, for example, the system slowed significantly, eventually requiring a shutdown to prevent a crash, which resulted in an unusually large, corrupted snapshot (9.3 GB) that exceeded the system's 8 GB RAM capacity.

\subsection{When to Use: Memory Analysis vs. Network Packet Analysis}
For a MySQL single-server setup, memory analysis is effective and fast for examining decrypted data directly from in-process memory. By capturing data in its decrypted state without needing to process encryption layers or reassemble network packets, memory analysis allows investigators to quickly access relevant information, making it ideal for time-sensitive forensic tasks.  However, when the goal is to monitor how MySQL queries and responses travel between the server and various clients or to identify patterns such as repeated access attempts from external IPs, network packet analysis is preferable. This method enables investigators to observe each query and response across the network, making it easier to detect unauthorized access attempts, excessive querying, or distributed attack patterns targeting the MySQL server. Together, both Memory Analysis and Network Packet Analysis offer a comprehensive forensic solution.

\section{Conclusion}

Conventional database logs, as demonstrated in the threat model, are fundamentally unreliable as a sole source of forensic evidence. A privileged attacker can disable, alter, or purge these records, creating significant blind spots for investigators. This paper confronted this challenge directly by introducing MemTraceDB, a tool that bypasses compromised logs entirely to reconstruct user activity from the ground-truth evidence source of process memory. This work provides a systematic and repeatable methodology for carving and interpreting volatile forensic artifacts from a running MySQL process, proving that a rich history of user actions persists in memory even when disk-based records have been destroyed.

The experiments presented in this paper validated this approach and yielded a critical empirical finding: the MySQL query stack has a finite operational capacity of approximately 9,997 queries. This discovery allowed for the establishment of a practical formula for determining snapshot frequency, providing investigators with a clear, actionable guideline for evidence collection. While the tests also identified limitations related to scalability and attribution under heavy load, they successfully proved the viability of the core methodology in typical single- and multi-user environments.

The development of MemTraceDB is a critical first step toward a more advanced, correlational approach to database forensics. Future work will focus on extending this methodology to other database systems and integrating the timelines generated by MemTraceDB with other evidence sources. By synthesizing disparate data streams, such as artifacts carved from persistent storage and application-level audit logs, a future correlational framework will enable the detection of sophisticated threats that would be invisible to any single source of analysis. Ultimately, this research provides a foundational technique for holding actors accountable in increasingly complex digital environments, paving the way for a new generation of intelligent, multi-source forensic systems.

\vspace{-3mm}
\bibliography{mybibfile}

\end{document}